# A New Prediction of Daily Number of New Cases and Total Number Infected for nCOVID-19 Plague Infections In Indonesia with the Modification of the Bernoulli Differential Equation

**Valentinus Galih Vidia Putra[1,a*], Juliany Ningsih Mohamad[2,b]**

[1] *Physics Department, Computational Physics Lab. Politeknik STTT Bandung, Bandung, 40272, Indonesia*

[2]*Physics Department, Computational Physics Lab.,* Faculty of Mathematics and Natural Sciences,*Universitas Nusa Cendana, Kupang, 85228, Indonesia*

e-mail: [a*]*valentinus@kemenperin.go.id*,
[b]ning_she@staf.undana.ac.id,
**\* Corresponding Author**

## Abstract

*The application of differential equations is commonly used in mathematics and physics, as well as various other sciences to explain a phenomenon in a system. This paper explains the mathematical modeling in the analysis of the nCOVID-19 plague in Indonesia on March 3, 2020, to April 19, 2020, with the modification of the Bernoulli equation and the simulation by MATLAB. In this study, it can be concluded that it was found that the daily number of nCOVID-19 cases in Indonesia will have the highest case at a maximum of around 400 and the total number of positive nCOVID-19 in Indonesia will reach 12000 people with a quiet period in mid-June. In this modeling, it has also been found that the value of R2 = 0.9927 on the total number of positive nCOVID-19 in Indonesia taken from 3 March 2020 to 19 April 2020, while the value of R2 = 0.807 daily number of positive new cases of nCOVID-19 in Indonesia taken from March 3, 2020, to April 19, 2020. Based on this research, it can be shown that the nCOVID-19 model for a case in the Indonesia plague is quite accurately compared by the real data.*

# Suatu Prediksi Baru Laju Harian Kasus Baru dan Jumlah Total Terinfeksi pada Kasus Infeksi Wabah nCOVID-19 di Indonesia dengan Modifikasi Persamaan Differensial Bernoulli

## Abstrak

Penerapan persamaan differensial umumnya digunakan pada ilmu matematika dan ilmu fisika serta berbagai ilmu lain untuk menjelaskan suatu fenomena pada sebuah sistem. Paper ini menjelaskan pemodelan matematik pada analisa wabah COVID-19 di Indonesia pada tahun 2020 pada 3 Maret 2020 hingga 19 April 2020 dengan modifikasi persamaan Bernoulli dan simulasi dengan MATLAB. Pada penelitian ini dapat disimpulkan bahwa jumlah pertambahan orang positif COVID-19 per hari di Indonesia akan mencampai puncak di sekitar titik maksimum 400 dan jumlah total orang yang terinfeksi COVID-19 di Indoensia akan mencapai 12000 orang dengan masa akhir pandemic sekitar pertengahan juni. Pada pemodelan ini juga telah divalidasi dengan data real dan ditemukan bahwa nilai $R^2$ = 0.9927 untuk jumlah total orang yang positif COVID-19 di Indonesia pada 3 Maret 2020 hingga 19 April 2020, sementara pertambahan harian kasus baru memiliki nilai $R^2$ = 0.807. berdasarkan hasil penelitian ini didapatkan bahwa hasil pemodelan memeprlihatkan hasil yang cukup baik untuk analisa kasus COVID-19 di Indoensia

**Kata Kunci:** Bernoulli, nCOVID-19, wabah di Indonesia, Fisika komputasi

# INTRODUCTION

The application of differential equations is commonly used in mathematics and physics, as well as various other sciences to explain a phenomenon in a system. Commonly, an ordinary differential equation is used in mathematics, physics, engineering and various other sciences to explain a phenomenon in a system [1-17]. In mathematics and in physics, the topic theory of the ordinary differential equations can be found in various fields of physics. Ordinary differential equation is one of the equations that exists as an independent variable, for example, time (*t*) and there are one or more derivatives as an independent function variable, for example, a derivatives $I(t) = \frac{dI(t)}{dt}$. In a model with the ordinary differential equation, the value of variable $I(t)$ is very important to obtain and to determine accurately. The results of the value of $I(t)$ can generally be used to explain the phenomenon in some cases [1]. Ordinary differential equations can be found in many fields of physics, such as geometry, mechanics, astronomy, particle physics, medical physics, textile physics, material physics and so on said some researchers such as Brooks-Pollock, E. and K.T.D. Eames [2], Blanchard, P. Robert, LD, and Glen, RH [1], Grassly NC, Fraser C. [3], Stone, L., Shulgin, B., Agur, Z [4], Rihan, FA & Anwar, NM [5], Nuraini, Khairudin, & Apri [6], Hyman & Li [7]. The application of the model using the system of differential equations in the textile field is carried out by several researchers such as Putra, Maruto & Rosyid, [8] and Putra, Rosyid & Maruto [9]. One of the applications of the ordinary differential equations in the medical field is modeling to analyze the spread of epidemic infections. Some forms of the ordinary differential equations in analyzing the spread of the plague can be shown in Equation (1), Equation (2) and Equation (3) [7]

$$\frac{dS}{dt} = \mu(S^0 - S) - \lambda S \qquad (1)$$

$$\frac{dI}{dt} = \lambda S - (\mu + \gamma + \delta)I \qquad (2)$$

$$\frac{dR}{dt} = \gamma I - (\mu + \xi)R \qquad (3)$$

The parameters $S, I, R$ are individuals who are susceptible affected by plague, infected individuals and recovered individuals. The parameters $\mu, \gamma, \delta, \xi, \lambda$ are constants that affect the rate of parameters S, I, R. Some researchers such as Brooks-Pollock, E. & K.T.D. Eames [2], Blanchard, P. Robert, LD, & Glen, RH [1], Stone, L., Shulgin, B., Agur, Z [3], Rihan, FA & Anwar, N. M [5], Hyman & Li [7] using SIR modeling, namely the general equation of the model is as in Equation (4), Equation (5) and Equation (6)

$$\frac{dS}{dt} = -\alpha SI \qquad (4)$$

$$\frac{dI}{dt} = \alpha SI - \beta I \qquad (5)$$

$$\frac{dR}{dt} = \beta I \qquad (6)$$

With parameters S, I, R are susceptible individuals affected by epidemics, infected individuals and recovered individuals. Values α, β are constants that affect the rate of parameters S, I, R. The researchers modeled the accurate model of the SIR equation to explain the optimum point of spread of plague infections, predict the quite period of plagues and also determine the number of individuals susceptible to plagues. A type of model of a plague of a virus especially nCOVID-19 has also been developed by Nuraini, Khairudin, Apri [6] using the Richard Curve by modification the Logistics curve model as in Equation (7)

$$\frac{dy(t)}{dt} = \frac{r}{\alpha} y \left(1 - \left(\frac{y}{K}\right)^{\alpha}\right) \text{ dan } y(t) = \frac{K}{(1 + \alpha e^{(-r\Delta t)})^{1/a}} \quad (7)$$

Where y (t) is a cumulative contaminated case, K is the maximum confirmed predictive case, α, r is the data fitting coefficient, Δt is the difference in days since the first time the case. The results of modeling Nuraini, Khairudin, Apri, [6] use the form of logistic model equations. Logistic models are generally used in cases to predict an epidemiology. This model is usually used to analyze risk factors for an plague and predict the possibility of an plague based on risk factors. A case of acute pneumonia caused by the nCOVID-19 virus in China has also been investigated by Lin Jia, Kewen Li, Yu Jiang, Xin Guo & Ting zhao [10] by using modeling through the incorporation of several models such as the logistic model, the Bertalanffy model and also the Gompertz model , the predicted results on the model are accurate enough to predict the spread of the nCOVID-19 plague in Wuhan, China. Some models of plagues of a virus have also been carried out by some researchers [10-17]. Some researchers such as Lipsitch M, Finelli L, Heffernan RT, Leung GM, Redd SC [14], Eubank S, Guclu H, Kumar VSA, Marathe M, Srinivasan A, et al. [15], Wallinga J, Teunist P [16], Barrett C, Bisset K, Leidig J, Marathe A, Marathe M [17] Grassly NC [3], Keeling MJ, Spiritual P [11], Yuan DF, Ying LY , Dong CZ [12], Zhang F, Li L, Xuan H Y. [13], Lipsitch M, Finelli L, Heffernan RT, Leung GM, Redd SC [14], Lin Jia, Kewen Li, Yu Jiang, Xin Guo & Ting zhao [10] used like the following models: Logistics Model is a model used in epidemiology and is used to analyze risk factors. The form of Logistics model is, like Equation (8)

$$Q(t) = \frac{a}{1 + e^{(b-c)(\Delta t)}} \quad (8)$$

Where Q (t) is a cumulative case of contamination, *a* maximum of confirmed case predictions, *b, c* is the data fitting coefficient, Δt is the difference in days since the first time of the case. The Bertalanffy model is a model commonly used to analyze the magnitude of the plague of the infection and also the growth of the population. The development of the plague infection is similar to the population and individual growth. The Bertalanffy model is usually also used to model the spread of the infections. The form Bertalanffy model can be shown as in Equation (9)

$$Q(t) = a(1 - e^{-b(\Delta t)})^c \quad (9)$$

Where Q (t) is a cumulative case of contamination, *a* maximum of confirmed case predictions, *b, c* is the data fitting coefficient, Δt is the difference in days since the first time the case. The Gomperts Model is a model used to predict the development population of an animal. This model can be used to explain the spread of an plague infection and study the factors that control and influence the spread of an plague infection. The form of this model can be shown in Equation (10)

$$Q(t) = ae^{-be^{-c(\Delta t)}} \quad (10)$$

Where Q (t) is a cumulative case of contamination, *a* maximum of confirmed case predictions, *b, c* is the data fitting coefficient, Δt is the difference in days since the first time the case. In modeling nCOVID-19 cases in Indonesia, the results of Nuraini, Khairudin, Apri, [6] model uses a modified form of logistic model equation to be able to predict the prediction number of nCOVID-19 cases in Indonesia and the number of new cases of nCOVID-19 in Indonesia by following the curve model prediction in South Korea. The modeling results can be shown in Figure 1. Nuraini, Khairudin, Apri, [6] modeling results show that starting the nCOVID-19 epidemic in Indonesia was at the beginning of March 2020 and the peak of the epidemic at the end of March 2020 and will end in mid-April 2020 with a number of cases a maximum of 8000 cases and an increase in the number of cases per day at 600 at the end of March.

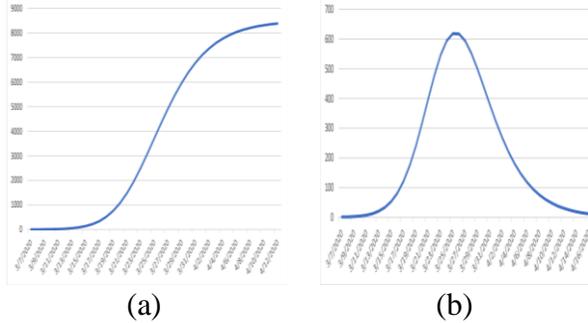

(a)          (b)

**Figure 1. a) projection of the number of nCOVID-19 cases in Indonesia; b) the number of new cases of nCOVID-19 in Indonesia [6]**

In this study, a pattern of the spread of nCOVID-19 cases in Indonesia was examined by applying the application of computational physics using MATLAB / Simulink in analyzing and completing the modeling of nCOVID-19 plague infections using the modification of the Bernoulli equation as in Equation (11) below

$$\frac{dI}{dt} + pI = qI^n \qquad (11)$$

## I. RESEARCH METHOD

In this modeling the Bernoulli equation model is used in predicting nCOVID-19 cases in Indonesia and in making a new scheme of MATLAB/simulink. Case data is taken from March 3, 2020 to April 19, 2020. The real data is taken from the data on the total number of patients infected with nCOVID-19 and also the rate of increase of nCOVID-19 patients every day. The limitation of this model is that the situation applied in Indonesia is strict social distancing and the public is very aware of the importance of social distancing and there is no large-scale migration of people from abroad into the country. In the Bernoulli model as in Equation (11) can be solved in the following way: Suppose that $I = y^{1-n}$, then it can be written a time differential like Equation (12)

$$\frac{dI}{dt} = (1-n)y^{-n}y' \qquad (12)$$

Multiply the two parts in Equation (11) with $(1-n)y^{-n}y'$ hence we get Equation (13)

$$(1-n)y^{-n}y' + (1-n)y^{-n}py = q(1-n)y^n y^{-n} \qquad (13.a)$$

Substitute Equation (12) to Equation (13.a), so that Equation (13.b) and Equation (13.c) are obtained.

$$\frac{dI}{dt} + (1-n)pI = (1-n)q \qquad (13.b)$$

$$\frac{dI}{dt} + GI = N \qquad (13.c)$$

The results of solving equation (13.c) can be described as follows Equation (14)

$$(D + G)I = N \qquad (14)$$

It can be assumed that the value of N = 0, so that Equation (15) and Equation (16) are obtained

$$(D + G)I = 0 \qquad (15)$$

$$I = A \exp\left(-\int G\, dt\right)$$
$$= A \exp(-Gt) \qquad (16)$$

Substitute Equation (16) to Equation (14) then get Equation (17) to Equation (20)

$$\frac{dA\exp(-Gt)}{dt} + GA\exp(-Gt) = N \qquad (17)$$

$$A\frac{d\exp(-Gt)}{dt} + \exp(-Gt)\frac{dA}{dt} + GA\exp(-Gt) = N \qquad (18)$$

$$-AG\exp(-Gt)$$
$$+\exp(-Gt)\frac{dA}{dt}$$
$$+GA\exp(-Gt) = N \tag{19}$$
$$\exp(-Gt)\frac{dA}{dt} = N \tag{20}$$

So we get equation (21)

$$A = \int N\exp(Gt)\,dt \tag{21}$$

With the value of I (t) being the same as Equation (22)

$$I(t) = \left[\int N\exp(Gt)\,dt + C\right]\exp(-Gt)$$
$$= \left[\int((1-n)q)\exp([(1-n)p]t)\,dt + C\right]\exp(-[(1-n)p]t) \tag{22}$$

In modeling the Plague, the following equation can be modified (Equation (11) as shown in Equation (23) to Equation (26))

$$\frac{dI}{dt} + pI = qI^n \tag{23}$$

$$\frac{dI}{dt} = pI - qI^n = p\left(I - \frac{q}{p}I^n\right) \tag{24}$$

if $n = 1 + m$, hence we get

$$\frac{dI}{dt} = p\left(I - \frac{q}{p}I^{(1+m)}\right)$$
$$= p\left(I - \frac{q}{p}I^{(1+m)}\right) \tag{25}$$
$$= pI\left(1 - \frac{q}{p}I^m\right)$$

if $p = \alpha = \frac{r}{\Omega}$ and $q = \beta p$ and $\frac{1}{\theta^m} = \beta$

$$\frac{dI}{dt} = \alpha I(1 - \beta I^m)$$
$$= \frac{r}{\Omega}I\left(1 - \left(\frac{I}{\theta}\right)^m\right) \tag{26}$$

Equation (26) has a solution that is Equation (27) and Equation (28)

$$I(t) = \left[\int((1-n)q)\exp([(1-n)p]t)\,dt + C\right]\exp(-[(1-n)p]t)$$
$$= \left[\int(m\beta\alpha)\exp(m\alpha t)\,dt + C\right]\exp(-m\alpha t) \tag{27}$$

$$I(t) = \left[\int\left(m\frac{r}{\Omega\theta^m}\right)\exp\left(m\frac{r}{\Omega}t\right)\,dt + C\right]\exp\left(-m\frac{r}{\Omega}t\right) \tag{28}$$

Where r, Ω, θ and m are fitting constants that will be adjusted to the real data of the number of nCOVID-19 cases in Indonesia and the number of new cases of nCOVID-19 in Indonesia. The modeling of total plague infections and the rate per day can be completed using MATLAB / Simulink in Equation (26) with the model shown in Figure 2 (r = 0.0992, Ω = 0.376, θ = 12000, m = 0.376)

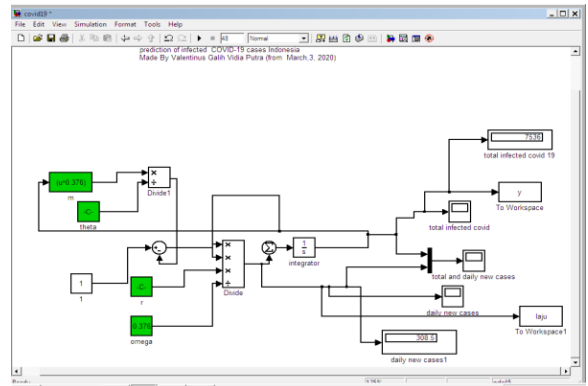

**Figure 2: MATLAB / Simulink Equation Modeling (26)**

Simulation results can be shown as follows (Figure 3 and Figure 4) by filling in the values of constants and performing computational and simulation calculations, then it is obtained. The prediction curve is the daily number of new nCOVID-19 cases and the total number of nCOVID-19 cases in Indonesia.

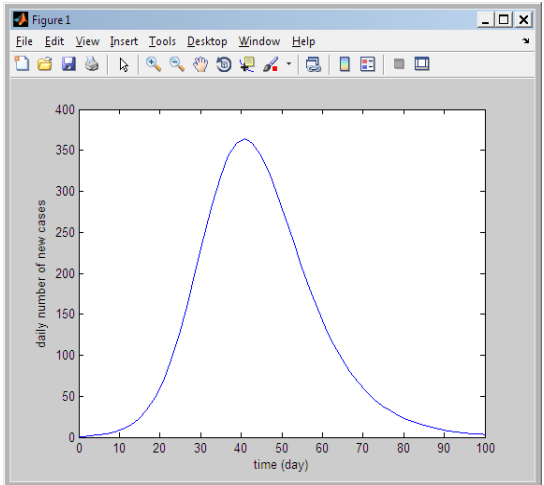

**Figure 3 daily number of new cases of nCOVID-19 in Indonesia**

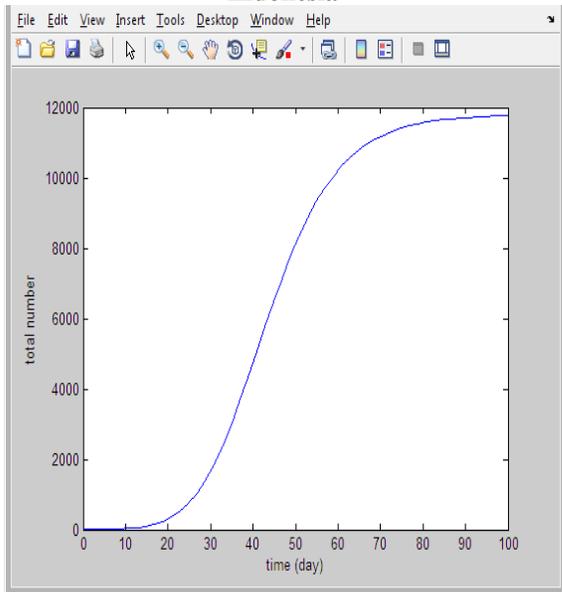

**Figure 4 Total number of COVID-19 cases in Indonesia**

Based on the results of modeling and with real data taken from March 3, 2020 to April 19, 2020, it can be analyzed the suitability of the model with real data [18] as in Figure 5 and Figure 6 the daily number of new cases and the total positive nCOVID-19 in Indonesia.

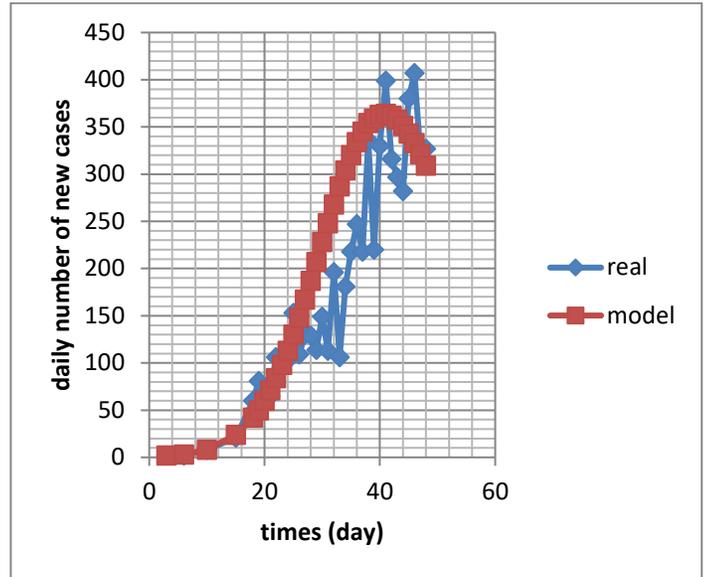

**Figure 5 Daily number of new cases of nCOVID-19 in Indonesia real data and predictions**

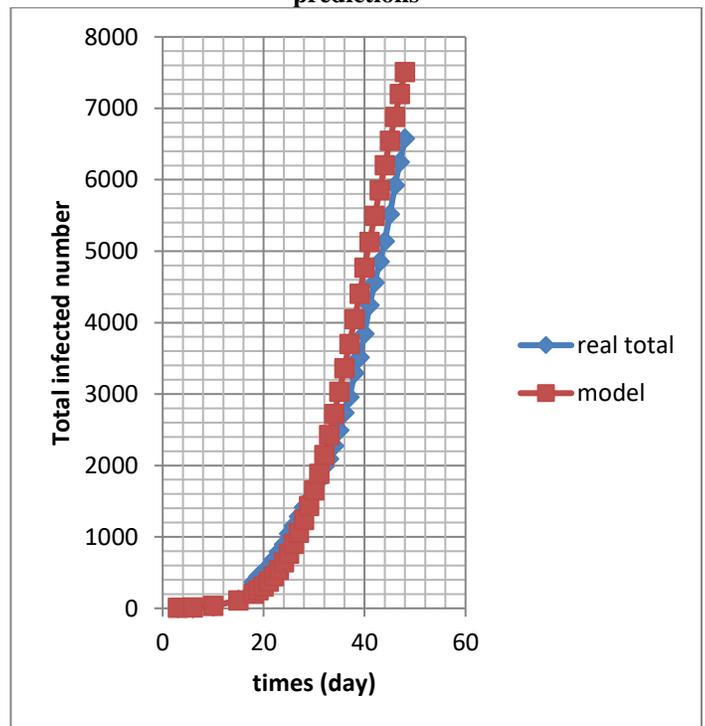

**Figure 6. Total number of positive cases of COVID-19 in Indonesian Prediction and Real Data**

It can be shown in Table 1. below regarding real and predicted data along with the value of $R^2$ the daily number of new cases and the total positive nCOVID-19 in Indonesia taken from March 3, 2020 to April 19, 2020.

Table 1. real and predicted data with $R^2$ daily number of new cases and positive total nCOVID-19 [18]

$R^2$ for case rates per day 0.807

| days per March,3, 2020 | Total real positive n COVID-19 | Rate per day | Rate per day prediction | Total nCOVID-19 prediction |
|---|---|---|---|---|
| 3 | 2 | 2 | 2 | 6 |
| 6 | 4 | 2 | 3 | 13 |
| 10 | 27 | 8 | 8 | 35 |
| 15 | 117 | 21 | 24 | 108 |
| 18 | 369 | 60 | 42 | 203 |
| 19 | 450 | 81 | 50 | 250 |
| 20 | 514 | 64 | 60 | 305 |
| 21 | 579 | 65 | 71 | 371 |
| 22 | 685 | 106 | 84 | 448 |
| 23 | 790 | 105 | 98 | 538 |
| 24 | 893 | 103 | 113 | 643 |
| 25 | 1046 | 153 | 130 | 764 |
| 26 | 1155 | 109 | 148 | 903 |
| 27 | 1285 | 130 | 167 | 1060 |
| 28 | 1414 | 129 | 187 | 1237 |
| 29 | 1528 | 114 | 207 | 1433 |
| 30 | 1677 | 149 | 228 | 1651 |
| 31 | 1790 | 113 | 248 | 1884 |
| 32 | 1986 | 196 | 268 | 2147 |
| 33 | 2092 | 106 | 287 | 2424 |
| 34 | 2273 | 181 | 304 | 2720 |
| 35 | 2491 | 218 | 320 | 3032 |
| 36 | 2738 | 247 | 334 | 3359 |
| 37 | 2956 | 218 | 345 | 3699 |
| 38 | 3292 | 336 | 354 | 4048 |
| 39 | 3512 | 220 | 359 | 4405 |
| 40 | 3842 | 330 | 363 | 4767 |
| 41 | 4241 | 399 | 364 | 5131 |
| 42 | 4557 | 316 | 362 | 5493 |
| 43 | 4854 | 297 | 358 | 5851 |
| 44 | 5136 | 282 | 351 | 6202 |
| 45 | 5516 | 380 | 343 | 6545 |
| 46 | 5923 | 407 | 333 | 6878 |
| 47 | 6248 | 325 | 321 | 7199 |
| 48 | 6575 | 327 | 309 | 7508 |

$R^2$ for total cases 0.9927

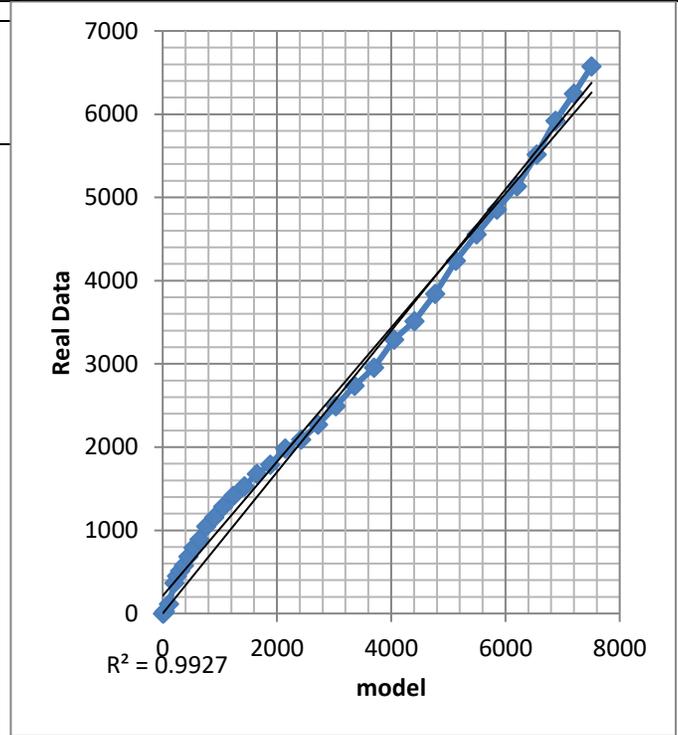

Figure 7. $R^2$ values The total number of positive nCOVID-19 in Indonesia taken from March 3, 2020 to April 19, 2020

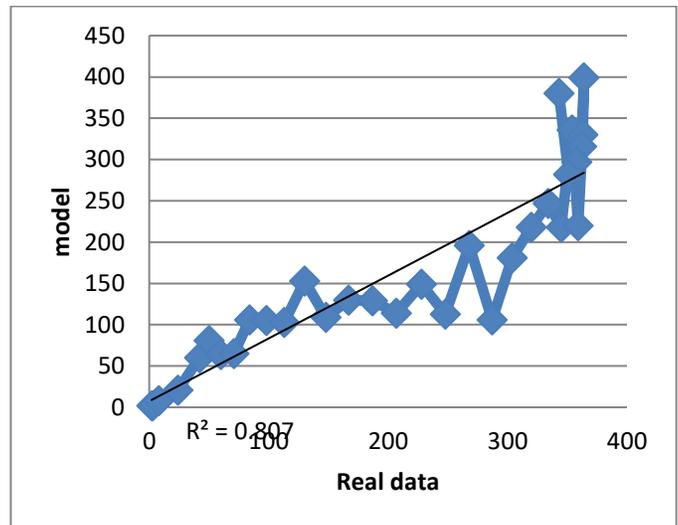

Figure 8 $R^2$ values of daily nCOVID-19 positive new cases in Indonesia taken from March 3, 2020 to April 19, 2020

The results of the simulation form $\frac{dI}{dt} = \frac{r}{\Omega}I\left(1-\left(\frac{I}{\theta}\right)^m\right)$ with MATLAB / Simulink can be shown as follows (Table-2)

**Table 2. Modeling predictions with several input parameters**

| Kind of models | $r$ | $\Omega$ | $\theta$ | $m$ |
|---|---|---|---|---|
| Model-1 | 0.0992 | 0.376 | 12000 | 0.376 |
| Model-2 | 0.15 | 0.376 | 12000 | 0.376 |
| Model-3 | 0.2 | 0.376 | 12000 | 0.376 |
| Model-4 | 0.0992 | 0.376 | 14000 | 0.376 |
| Model-5 | 0.0992 | 0.376 | 18000 | 0.376 |
| Model-6 | 0.0992 | 0.8 | 12000 | 0.8 |
| Model-7 | 0.0992 | 1.2 | 12000 | 1.2 |
| Model-8 | 0.0992 | 2 | 12000 | 1.2 |
| Model-9 | 0.0992 | 1.2 | 12000 | 2 |

The simulation results in Table-2 can be shown in Figure 9 to Figure 17

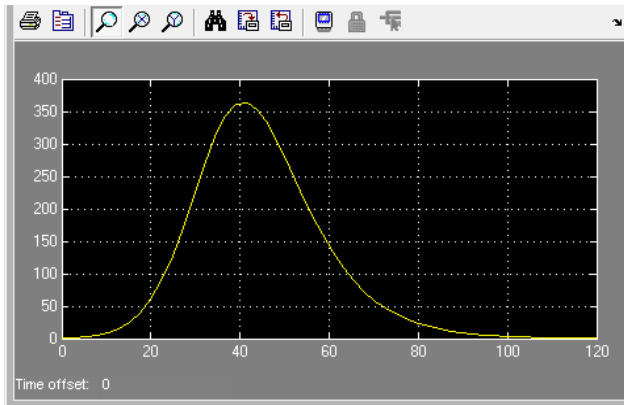

**Figure 9 Results of modeling with model-1**

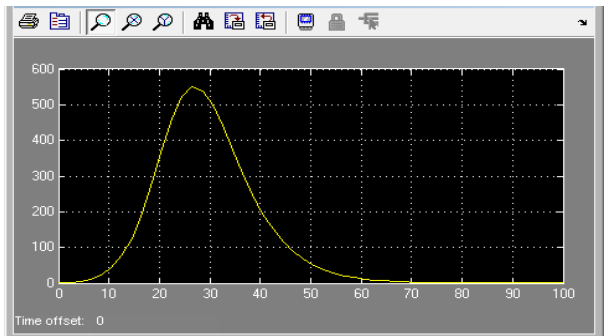

**Figure 10 Results of modeling with model-2**

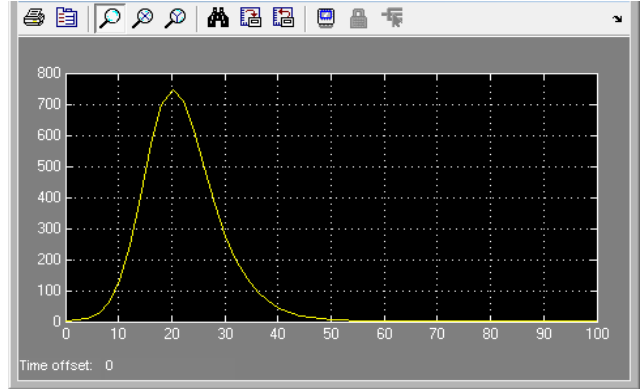

**Figure 11 Results of modeling with model-3**

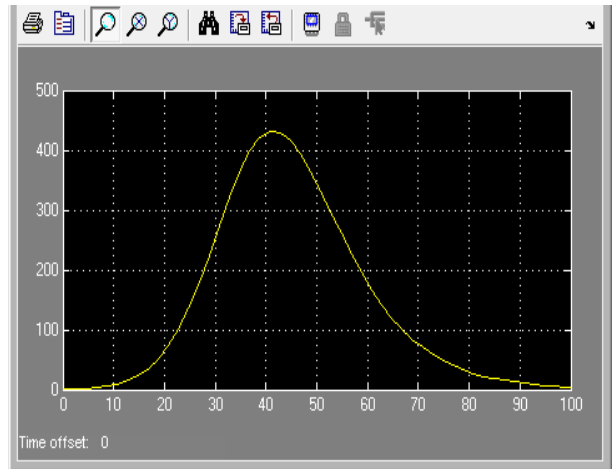

**Figure 12 Results of modeling with model-4**

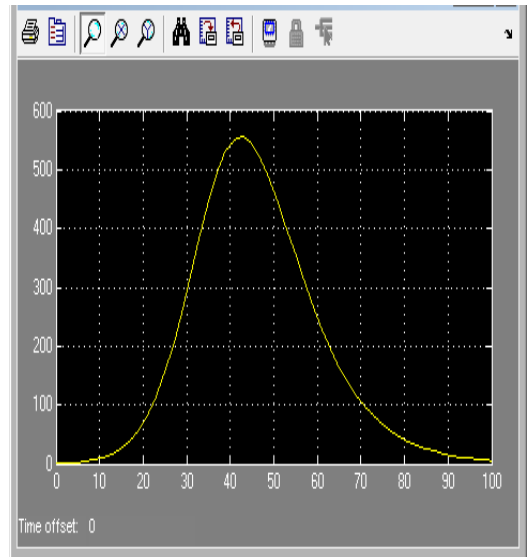

**Figure 13 Results of modeling with model-5**

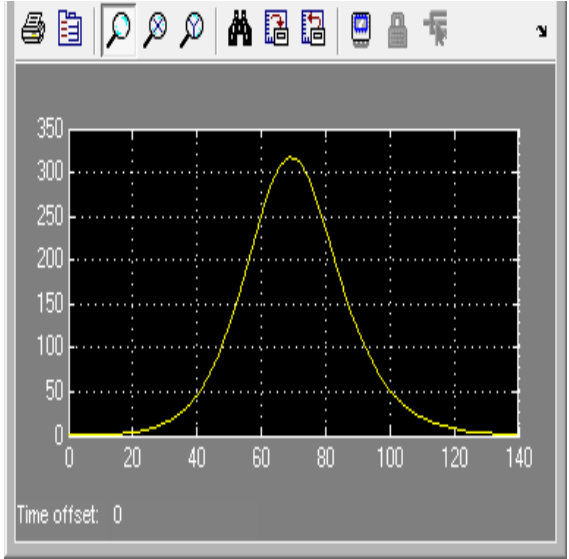

**Figure 14 Results of modeling with model-6**

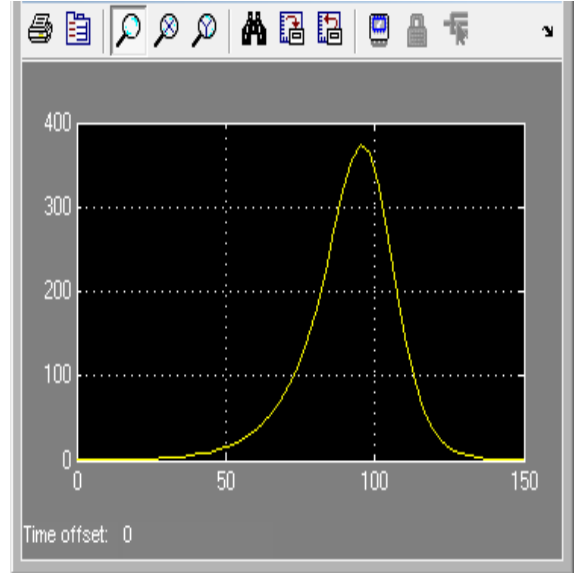

**Figure 17 Results of modeling with model-9**

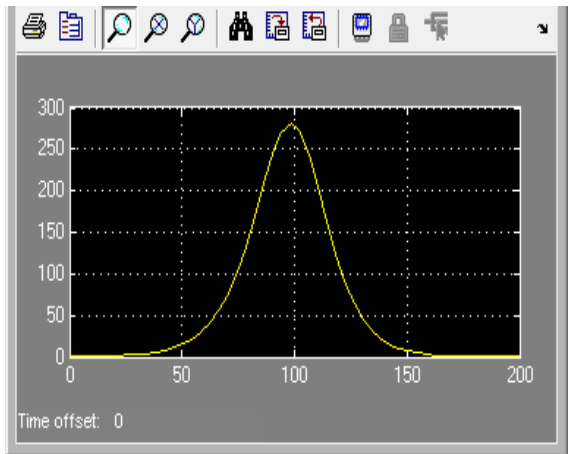

**Figure 15 Results of modeling with model-7**

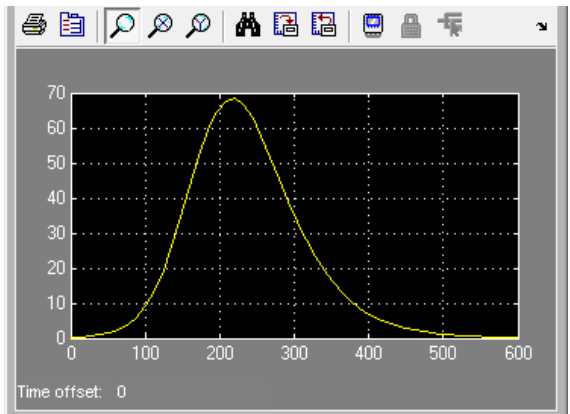

**Figure 16 Results of modeling with model-8**

Based on the simulation results, it is found that for the greater *r* value while other parameters are constant (such as model-1 to model-3), the shape of the speed curve per day will be more acute with a relatively shorter time but a higher peak speed per day. Based on the simulation results it also found that for the greater value θ (such as model-4 and model-5), then the peak point of the rate curve per day will be even greater with the same duration of time. Based on the simulation results also found that for the value $\Omega = m$ (such as model-6 and model-7) the greater the peak point of the infected people, the rate curve per day will be smaller but have a longer duration of time. For a case when $\Omega \neq m$ (such as model-8 and model-9), the $\Omega$ will influence the daily number of new cases with a lower people infected by the virus but longer time to end the plague, as well as the *m* will influence the daily number of new cases with higher people infected by the virus but in shorter time of period.

## II. RESULTS AND DISCUSSIONS

In this modeling, a modified model of the Bernoulli equation has been examined in predicting nCOVID-19 cases in Indonesia.

Case data is taken from March 3, 2020, to April 19, 2020. The data taken is the real data on the total number of patients infected with nCOVID-19 and also the rate of increase of nCOVID-19 patients every day. From the results of the equation, it was found that the modified model of the Bernoulli equation can be used to predict the daily number of new cases and the total positive nCOVID-19 in Indonesia. The modeling results used in this equation are as in Equation (29)

$$\frac{dI}{dt} = \alpha I(1 - \beta I^m)$$
$$= \frac{r}{\Omega} I\left(1 - \left(\frac{I}{\theta}\right)^m\right) \quad (29)$$

With the solution of the results of the equation which is like Equation (30)

$$I(t) = \left[\int \left(m\frac{r}{\Omega\theta^m}\right) \exp\left(m\frac{r}{\Omega}t\right) dt + C\right] \exp(-m\frac{r}{\Omega}t) \quad (30)$$

For a case $\Omega = m$, then the form of Equation (29) will have a shape that is almost similar to the plague model of a nCOVID-19 virus which has also been developed by Nuraini, Khairudin, Apri [6] by using the Richard Curve Logistics development model as in Equation (7), so that the Equation form (29) can be modified to Equation (31) below

$$\frac{dI(t)}{dt} = \frac{r}{\Omega} y \left(1 - \left(\frac{y}{\theta}\right)^\Omega\right) \quad (31)$$

Where $\Omega$, r and $\theta$ are fitting constants. Through modeling using MATLAB / Simulink, it was found that the daily number of nCOVID-19 cases in Indonesia will have the highest case at a maximum of around 400 and the total number of positive nCOVID-19 in Indonesia will reach 12000 people with a quiet period in mid-June. In this modeling it has also been found that the large value of $R^2 = 0.9927$ on the total number of positive COVID-19 in Indonesia taken from 3 March 2020 to 19 April 2020, while the value of $R^2 = 0.807$ daily number of positive new cases of COVID-19 in Indonesia taken from March 3, 2020 to April 19, 2020. Based on the study, the situation applied in Indonesia with strict social distancing and the awareness of the community seriously about the importance of social distancing and there is no movement of people from abroad into the country on a large scale will alleviate the plague of nCOVID -19 this in mid-June. In this study, it has also been found that MATLAB / Simulink in Figure 2 can be used to solve the case of a differential equation with the Bernoulli equation in the nCOVID-19 plague modeling example with sufficient accuracy. Based on the simulation results, it is found that for the greater r-value while the other parameters are constant, the shape of the speed curve per day will be more acute with a relatively shorter time but a higher peak speed per day. Based on the simulation results also found that for the greater value θ, then the peak point of the rate curve per day is also greater with the same duration of time. Based on the simulation results also found that for the value of $\Omega = m$ the greater, then the peak point of the rate curve per day will be smaller but has a longer time duration. At the very beginning of the COVID-19 plague, it was believed to be a disease transmitted by respiratory droplets through close person-to-person contact. Respiratory droplets are relatively large-sized particles and thus cannot travel long distances through air and therefore this transmission cannot account for the rapid and widespread disease. The increase of the virus spread can be stopped by the social distancing method. In handling the nCOVID-19 pandemic problem in Indonesia, it can be shown that the Indonesian government has done a pandemic reduction that can be demonstrated with the model 1 curve. Curve model 1 shows that the nCOVID-19 pandemic will last approximately three months but at a daily rate of nCOVID-19 patients which is quite low. Model-6 and model 7 can minimize the rate per day of nCOVID-19 patients but have an exposed period of around five months of a pandemic. Based on this model it can be shown that the government must be able to make a choice regarding the length of the pandemic which must be as minimal as possible but with a minimal number of rates per day for COVID-19 patients. With the methods currently carried out by the government, such as social distancing, calls for health care and public awareness, it can be predicted that the length of the pandemic will last for three months.

### III. CONCLUSIONS

In this study, it can be concluded that it was found that the daily number of nCOVID-19 cases in Indonesia will have the highest case at a maximum of around 400 and the total number of positive nCOVID-19 in Indonesia will reach 12000 people with a quiet period in mid-June. In this modelling, it has also been found that the value of $R^2 = 0.9927$ on the total number of positive nCOVID-19 in Indonesia taken from 3 March 2020 to April 19, 2020, while the value of $R^2 = 0.807$ daily number of positive new cases of nCOVID-19 in Indonesia taken from March 3, 2020 to April 19, 2020. Based on this research, it can be shown that the nCOVID-19 model for a case in Indonesia plague is quite accurately compared by the real data.

### ACKNOWLEDGMENT

The author thanks to the Department of physics Politeknik STTT Bandung and Nusa Cendana University who have supported this research activity so that this research can be completed well.

### REFERENCES

[1] Blanchard, P. Robert, L.D., & Glen,R.H. *Ordinary differential Equations*, Richard Stratton, 2012.
[2] Brooks Pollock, Ellen & Eames, Ken T.D..Pigs didn't fly, but swine flu:, *Mathematics Today*, 2011, vol.47,Hal..36-40.
[3] N C Grassly，C. Fraser, Mathematical models of infectious disease transmission．*Nature Reviews Microbiology*, 2008, 6(6):477-487
[4] L. Stone, B. Shulgin, Z.Agur, Theoretical Examination of the Pulse Vaccination Policy in the SIR Epidemic Model, Proceedings of the Conference on Dynamical Systems in Biology and Medicine), *Math. Comput. Modelling* , 2000,Vol. 31, No.4-5, 207-215.
[5] F. A. Rihan, & N. M. Anwar, Qualitative analysis of delayed SIR epidemic model with a saturated incidence rate, *International Journal of Differential Equations*, 2012,Art. pp. ID 408637, 13
[6] N. Nuraini, K. Khairudin,M Apri, *Data dan Simulasi COVID-19 dipandang dari Pendekatan Model Matematika*, Bandung: Institut Teknologi Bandung, *2020.*
[7] M. James Hyman &Jia Li , Differential susceptibility epidemic models, *J. Math. Biol*, *2005*. 50, 626–644*,*
[8] VGV Putra., G. Maruto & M.F Rosyid, New theoretical modeling for predicting yarn angle on OE yarn influenced by fibre movement on torus coordinate based on classical mechanics approach**,** *Indian Journal of Fibre and Textile Research*, Vol.42, , 2017 pp. 359-363.
[9] VGV Putra, M.F. Rosyid, & G.Maruto, ,A Simulation Model of Twist Influenced by Fibre Movement Inside Yarn on Solenoid Coordinate, *Global Journal of Pure and Applied Mathematics*, , 2016Vol 12.,No.1, pp. 405-412.

[10] Lin Jia, Kewen Li, Yu Jiang, Xin Guo & Ting Zhao, *Prediction and analysis of Corona virus Disease 2019*, Beijing: China university of Geosciences, 2020

[11] M J Keeling & P Rohani．*Modeling Infectious Diseases in Humans and Animals*．New Jersey：Princeton University Press，2007

[12] D F Yuan, L Y Ying & C Z Dong, Research Progress on Epidemic Early Warning Model. *Zhejiang Preventive Medicine*, , 2012 ,(08):20-24+27.

[13] F Zhang, L Li, H Y. Xuan Overview of infectious disease transmission models. *Theory and Practice of Systems Engineering*, 2011, 31(9):1736-1744.

[14] M Lipsitch, L Finelli, RT Heffernan, GM Leung, SC Redd, Improving the evidence base for decision making during a pandemic: the example of 2009 inuenza A/H1N1. *Biosecurity and bioterrorism biodefense strategy practice and science,* 2011 9: 89–115.

[15] S Eubank, H Guclu, VSA Kumar, M Marathe, A Srinivasan, et al. Modelling disease out breaks in realistic urban social networks. *Nature*, 2004, 429: 180–184.

[16] J Wallinga & P Teunis, Different epidemic curves for severe acute respiratory syndrome reveal similar impacts of control measures. *American Journal of Epidemiology* . 2004,160: 509–516.

[17] C Barrett, K Bisset, J Leidig,A Marathe,M Marathe., Economic and social impact of inuenza mitigation strategies by demographic class. *Epidemics*, 2011, 3: 19–31.

[18] https://www.kemkes.go.id/article/view/20031900002/Dashboard-Data-Kasus-COVID-19-di-Indonesia.html. accessed 19 April 2020